\documentclass[12pt,a4paper,fleqn]{article}
\usepackage[textheight=23cm,textwidth=16cm]{geometry}
\usepackage[utf8]{inputenc}
\usepackage{amsmath}
\usepackage{amsfonts}
\usepackage{graphicx}
\usepackage[superscript]{cite}

\usepackage[american]{babel}

\usepackage{dcolumn} 
\newcolumntype{d}{D{.}{.}{2}}
\newcolumntype{e}{D{.}{.}{3}}
\newcolumntype{f}{D{.}{.}{4}}
\usepackage{hyperref}

\usepackage{setspace}
\def\Bra#1{\left\langle#1\right|}
\def\Ket#1{\left|#1\right\rangle}

\newcommand{\rcm}{cm$^{-1}$}

\makeatletter
\def\@citess#1{\textsuperscript{[#1]}}
\makeatother

\begin{document}
\begin{center}

{\LARGE\bf
On the benefits of localized modes \\ in anharmonic vibrational calculations \\ for small molecules \\
}

\vspace{2cm}

{\large 
Pawe\l{} T. Panek,
Christoph R. Jacob\footnote{E-Mail: c.jacob@tu-braunschweig.de}
}\\[4ex]

TU Braunschweig, \\
Institute of Physical and Theoretical Chemistry, \\
Hans-Sommer-Str.~10, 38106 Braunschweig, Germany

\vfill

\end{center}

\begin{tabbing}
Date:   \qquad\qquad\qquad \= May 04, 2016 \\
Status:       \> \textit{J. Chem. Phys.} \textbf{144}, 164111 (2016).\\
DOI: \> \href{http://dx.doi.org/10.1063/1.4947213}{10.1063/1.4947213}
\end{tabbing}

\newpage

\begin{abstract}
Anharmonic vibrational calculations can already be computationally demanding for relatively small molecules. The main bottlenecks 
lie in the construction of the potential energy surface and in the size of the excitation space in the vibrational configuration interaction (VCI)
calculations. To address these challanges, we use localized-mode coordinates to construct potential energy surfaces and perform 
vibrational self-consistent field (\mbox{L-VSCF}) and \mbox{L-VCI} calculations [P.~T.~Panek, Ch.~R.~Jacob, \textit{ChemPhysChem} \textbf{15}, 3365 (2014)] 
for all vibrational modes of two prototypical test cases, the ethene and furan molecules. 
We find that the mutual coupling between modes is reduced when switching from normal-mode coordinates to localized-mode coordinates.
When using such localized-mode coordinates, we observe a faster convergence of the $n$-mode expansion of the potential energy surface. 
This makes it possible to neglect higher-order contributions in the $n$-mode expansion of the potential energy surface or to approximate 
higher-oder contributions in hybrid potential energy surfaces, which reduced the computational effort for the construction of the anharmonic
potential energy surface significantly. Moreover, we find that when using localized-mode coordinates, the convergence with respect to the 
VCI excitation space proceeds more smoothly and that the error at low orders is reduced significantly. This makes it possible to devise low-cost 
models for obtaining a first approximation of anharmonic corrections.
This demonstrates that the use of localized-mode coordinates can be beneficial already in anharmonic vibrational calculations of small
molecules, and provides a possible avenue for enabling such accurate calculations also for larger molecules.

\end{abstract}

\newpage

\section{Introduction}
\label{sec:Introduction}

Going beyond the harmonic approximation in theoretical vibrational spectroscopy
is an important step, as it gives access to chemically accurate spectroscopic and thermochemical data.
With some effort, for small-to-medium size molecules a quantitative agreement with experimental results 
can be achieved \cite{pfeiffer_anharmonic_2013,bloino_anharmonic_2015,tajti_heat:_2004}.
There are two major well-established approaches for calculations of anharmonic vibrational
spectra, vibrational perturbation theory (VPT) and variational methods. In the first one,
the anharmonicity is treated as a perturbation of the zeroth-order solution obtained in the harmonic
approximation \cite{barone_anharmonic_2005, cornaton_analytic_2016}. In the latter approach, on which
we will focus here, the nuclear Schr\"odinger equation is solved in a variational manner, analogously 
to electronic structure methods. Most prominent starting-point is the vibrational
self-consistent field (VSCF) method, on top of which the lacking vibrational correlation energy 
is approximated by methods such as vibrational configuration interaction (VCI) or (non-variational) vibrational
coupled cluster (VCC) theory \cite{christiansen_selected_2012, roy_vibrational_2013}.

However, all available approaches are feasible only for relatively small molecules. 
The first bottleneck lies in constructing the anharmonic potential energy surface (PES). 
Mapping the full PES directly for all vibrational coordinates is achievable only for molecules consisting 
of few atoms, as the number of required grid points grows exponentially with the number of nuclei. 
In VPT, the PES is thus approximated by a Taylor expansion of the energy 
at the equilibrium geometry, which is usually truncated at third or fourth order.
On the other hand, variational methods usually employ a truncated $n$-mode expansion \cite{jung_vibrational_1996,
carter_extensions_1998} of the PES, in which the potential energy is expanded discretely on a grid. 
An accurate approximation of the PES requires the inclusion of higher-order terms in the $n$-mode 
expansion. The main hindrance in the $n$-mode expansion is the need for the calculation
of coupling terms, i.e., of two-mode, three-mode, or even higher-order potentials.

Several methods have been developed to speed up the construction of the PES and in 
order to make anharmonic vibrational calculations feasible also for larger systems. 
For some combinations of modes, the couplings will be relatively small, which could be 
prescreened without a significant loss of accuracy \cite{benoit_fast_2003, pele_number_2008, 
rauhut_efficient_2004, yagi_multiresolution_2007, seidler_coupling_2009}.
Another possibility to tackle the computational cost of obtaining the higher-order terms
in the PES expansion are multi-level or hybrid approaches \cite{rauhut_efficient_2004,
hrenar_towards_2005, puzzarini_accurate_2010}, in which electronic-structure methods of different 
levels are applied to the different terms of the PES expansion. In the spirit of an hierarchical expansion, 
the most dominant one-mode terms are treated with high-level electronic-structure methods, such as 
coupled-cluster theory, whereas for the two-mode or higher-order terms less accurate but more 
efficient methods, such as density-functional theory, are employed. 

Since VSCF describes the interaction of modes only in a mean-field manner, post-VSCF methods such 
as VCI \cite{bowman_application_1979} have to be applied to recover the vibrational correlation energy. 
Here, the main computational bottleneck arises from the construction of the VCI-Hamiltonian matrix. Its 
size depends on the expansion of the VCI space. Converged VCI energies require rather large excitation 
spaces, which again limits the applicability of the method to small molecules.
This problem is addressed with several configuration selection methods, where only the mostly contributing 
states are included in the VCI expansion, vastly reducing the size of the Hamiltonian \cite{rauhut_configuration_2007,
scribano_iterative_2008, carbonniere_vci-p_2010,knig_automatic_2015}.

Thus, in general there are three main aspects that need to be addressed to make anharmonic vibrational calculations
feasible for larger molecules: 
(i) the convergence of the $n$-mode expansion and therein (ii) the possibility to neglect or approximate some
of the couplings between modes, and (iii) the convergence of the expansion of the excitation space in the post-VSCF
treatment, namely the VCI space. The choice of coordinates, in which the calculations are performed,
will affect all three of these aspects. Here, we will explore to what extent a suitable choice of the
coordinates can be beneficial for anharmonic vibrational calculations.
In the present study, we will focus on rectilinear coordinates. Curvilinear coordinates, though more natural, 
imply a more complicated form of the kinetic energy operator, which in turn makes the solution of the vibrational 
Schr\"odinger equation more demanding \cite{csaszar_exact_1995, yurchenko_theoretical_2007,
scribano_fast_2010,pesonen_normal_2012}.

Normal-modes coordinates, obtained in the harmonic approximation,
are the most common choice for anharmonic vibrational calculations. 
Unfortunately, in the anharmonic regime they provide a complicated picture of strong
couplings between the normal modes. This results in a slow convergence, both with respect 
to the $n$-mode expansion of the PES, and with respect to the VCI-expansion of the excitation 
space.

One way of defining more appropriate coordinates are so-called local modes \cite{henry_use_1977}.
They arise from the assumption that X--H stretching vibrations, instead of being collective 
motions of many X--H groups should rather be a set of coupled single-bond stretching motions,
and such strictly local vibrational coordinates are thus constructed.
Such local modes were successfully applied to O--H stretching vibrations of water clusters
and alcohols, giving access to overtone spectroscopy \cite{low_calculation_1999,
salmi_calculation_2009, howard_weak_2005}.
A special case of local modes is the local monomer model (LMon) presented by 
Bowman and coworkers \cite{wang_towards_2010,mancini_--flyab_2013,wang_coupled-monomers_2012}.
The FALCON scheme by König \textit{et al.} constructs strictly local vibrational coordinates 
in a similar spirit \cite{knig_falcon:_2016}. 

Another approach are optimized coordinates, which are obtained from a transformation of the normal modes coordinates 
during the VSCF procedure to minimize the VSCF energy \cite{thompson_optimization_1982,moiseyev_scf_1983}. 
Yagi and coworkers introduced optimized-coordinate variants 
of VSCF, VCI, and VCC and showed that in the new basis a faster convergence with respect to the excitation space
is observed.\cite{yagi_optimized_2012, thomsen_optimized_2014} These coordinates were also employed in a post-VSCF 
perturbative treatment \cite{yagi_vibrational_2014}. It was noticed that the optimization procedure can provide coordinates
that are well localized, and thus mutually decoupled. However, in this framework a full quartic force field is used
and transformed during the optimization, which precludes computational savings in the construction of the PES.

The locality of the coordinates can also be enforced during a transformation of normal modes coordinates. 
Such rigorously defined localized modes have been introduced for the analysis
and interpretation of selected bands of vibrational spectra of large molcules \cite{jacob_localizing_2009,
jacob_analysis_2009,jacob_understanding_2009,liegeois_analysis_2010,weymuth_local-mode_2010,jacob_theoretical_2011}.
We have previously developed localized-modes variants of VSCF and VCI, termed
\mbox{L-VSCF} and \mbox{L-VCI}, respectively \cite{panek_efficient_2014}. For selected vibrations of water clusters 
and polypeptides, we could show that localized-mode coordinates lead to an organized picture of the couplings between 
the localized modes, allowing for neglecting some of the coupling potentials \textit{a priori}, without a loss in accuracy. 
Moreover, a faster convergence within the VCI excitation space was observed.
Cheng and Steele applied such localized-modes coordinates only in the VSCF framework, and explored
using distance-based criteria for neglecting two-mode couplings \cite{cheng_efficient_2014}.
Hanson-Heine investigated the harmonic couplings arising from the localization
procedure, and proposed their utilization as a post-VSCF correlation correction in the \mbox{L-VSCF}(HC) method
\cite{hanson-heine_examining_2015}. Recently, Christiansen and co-workers proposed
hybrid optimized/localized vibrational coordinates, where both the energy and spatial localization
conditions are applied in the VSCF procedure \cite{klinting_hybrid_2015}.

In this paper, we explore the use of localized modes in the \mbox{L-VSCF}/\mbox{L-VCI} framework 
for calculations of the full anharmonic vibrational spectra (i.e., of all fundamental vibrations) for small molecules. 
With these coordinates, we assess their benefits with respect to the main bottlenecks of such computations,
the convergence of the $n$-mode expansion and the possibility to neglect its small contributions \textit{a priori}, 
as well as the convergence of the VCI expansion with respect to the excitation space.

This work is organized as follows: First, we recall and introduce the main aspects of the theory,
concerning the $n$-mode expansion (Section~\ref{sec:n-mode}), the localization of normal modes (Section~\ref{sec:Localization}),
and the L-VCSF and \mbox{L-VCI} methods (Section~\ref{sec:lvscf_lvci}). Subsequently, the computational details are described in 
Section~\ref{sec:comp}. Next, we apply these methods to perform anharmonic vibrational calculations for ethene in 
Section~\ref{sec:ethene} and for furan in Section~\ref{sec:furan}. With high-level anharmonic potential energy surfaces 
for these two test cases, we investigate the mutual couplings of normal and localized modes, respectively (Sections~\ref{sec:ethene_harmonic}
and~\ref{sec:furan_harmonic}), the convergence with respect to the VCI excitation space (Section~\ref{sec:ethene_vci}) and the 
convergence of the $n$-mode expansion (Sections~\ref{sec:ethene_nmode_ch-str}, 
\ref{sec:ethene_nmode_all}, and~\ref{sec:furan_potentials}). Our best-estimate (L-)VCI fundamental 
energies are compared to experimental reference data (Sections~\ref{sec:ethene_nmode_all} and~\ref{sec:furan_potentials}). 
At this point, we also explore simplified models to reduce the computational cost without a significant loss of accuracy. 
Finally, our conclusions are summarized in Section~\ref{sec:summary}. 

\section{$n$-mode expansion of the PES}
\label{sec:n-mode}

The potential energy surface in our calculations is approximated by the hierarchical
 \mbox{$n$-mode} expansion \cite{jung_vibrational_1996,carter_extensions_1998}, i.e.,
\begin{align}
  \label{eq:pes}
  V({\boldsymbol{q}}) =& \ \sum_i^M V_i^{(1)}({q}_i) + \sum_{i<j}^M V_{ij}^{(2)}({q}_i,{q}_j) \nonumber \\
  &+ \sum_{i<j<k}^M V_{ijk}^{(3)}({q}_i,{q}_j,{q}_k) +\sum_{i<j<k<l}^M V_{ijkl}^{(4)}({q}_i,{q}_j,{q}_k,{q}_l) + \ldots, 
\end{align}
with $\boldsymbol{q}$ being $M$ rectilinear coordinates,
\begin{align}
\label{eq:norm_coord}
{q}_i = \sum_{I=1}^{N_\text{nuc}} \sum_{\alpha=x,y,z} {Q}^i_{I\alpha} R_{I\alpha}^{(m)},
\end{align}
where $R_{I\alpha}^{(m)}$ is the mass-weighted Cartesian $\alpha$-coordinate ($\alpha = x,y,z$) of nucleus $I$ 
and where
\begin{align}
\label{eq:nm_coord_delta}
\sum_{I\alpha} Q_{I\alpha}^i Q_{I\alpha}^j = \delta_{ij}.
\end{align}
The coefficients ${Q}^i_{I\alpha}$ can be chosen as the normal-mode vectors obtained in the harmonic 
approximation, or as localized modes $\tilde{Q}^i_{I\alpha}$ obtained via a transformation of the normal 
modes. Here and in the following the tilde denotes quantities referring to localized modes. Details will 
be discussed in Section~\ref{sec:Localization}.

To make the calculation of the PES feasible, the $n$-mode expansion is truncated at a certain order. 
It should be stressed that the PES in a truncated $n$-mode expansion is not invariant 
upon transformation of the modes, and thus the PES in localized-mode coordinates is not equivalent 
to the PES in normal-mode coordinates. Therefore, the PES in different coordinates have to be constructed 
separately. Consequently, the vibrational frequencies obtained with a truncated $n$-mode expansion in 
normal-mode coordinates and in localized-mode coordinates, respectively, will not be identical. This difference 
will, however, vanish when the $n$-mode expansion is fully converged. In contrast, a truncated Taylor expansion 
of the PES can be easily transformed to a representation in other rectilinear modes (for details, see Appendix~A 
of Ref~\cite{yagi_optimized_2012}).

\section{Localization of normal modes}
\label{sec:Localization}

Here we recall only the crucial aspects of the method of normal modes localization, for further details
see Ref.~\cite{jacob_localizing_2009}. We begin with the normal modes $\boldsymbol{Q}$ obtained 
as eigenvectors of the mass-weighted molecular Hessian $\boldsymbol{H}^{(m)}$,
\begin{align}
  \boldsymbol{H}^{(q)}  = \boldsymbol{Q}^T \boldsymbol{H}^{(m)} \boldsymbol{Q},
\end{align}
with corresponding eigenvalues (i.e., squared vibrational frequencies) $H_{ii}^{(q)}=\omega_i^2=4\pi\nu_i^2$.
The corresponding normal-mode coordinates are then defined by Eq.~\ref{eq:norm_coord}.
To arrive at localized modes, the normal modes are transformed with a unitary transformation $\boldsymbol{U}$,
\begin{align}
\tilde{\boldsymbol{Q}} = \boldsymbol{Q} \boldsymbol{U},
\label{eq:Qloc}
\end{align}
such that $\boldsymbol{U}$ maximizes the localization measure $\xi(\tilde{\boldsymbol{Q}})$.
Here, we apply the atomic-contribution criterion defined as \cite{jacob_localizing_2009}
\begin{align}
\xi_{\rm at} \left (\tilde{\boldsymbol{Q}} \right) = \sum_{p=1} \sum_{i=1} \left ( \tilde{C}_{ip} \right)^2,
\label{eq:lm_pipek_mezey}
\end{align}
where $\tilde{C}_{ip}$ corresponds to the contribution of nucleus $i$ to the normal mode $\boldsymbol{Q}_p$
defined as
\begin{align}
\tilde{C}_{ip} = \sum_{\alpha=x,y,z} \left ( \tilde{Q}_{i\alpha,p} \right )^2.
\label{eq:lm_pipek_mezey2}
\end{align}
When localizing the normal modes, the Hessian in the basis of the localized modes,
\begin{align}
\label{eq:lm_hess}
\tilde{\boldsymbol{H}} = \boldsymbol{U}^T \boldsymbol{H}^{(q)} \boldsymbol{U},
\end{align}
is no longer diagonal and the localized modes are not the  eigenvectors of the molecular Hessian. 
%

Commonly, the localization of the normal modes is performed for subsets of normal modes. To this end, 
the matrix $\boldsymbol{Q}$ is divided into column vectors which are assigned to $n_s$ subsets such that
\begin{align}
\boldsymbol{Q} = \boldsymbol{Q}^{\rm sub,1} || \boldsymbol{Q}^{\rm sub,2} || \ldots || \boldsymbol{Q}^{\rm sub,n_s}.
\end{align}
This leads to $n_s$ distinct subsets of localized modes $\lbrace \tilde{\boldsymbol{Q}}^{\rm sub,i} \rbrace$. 
The resulting Hessian in the basis of these subset-localized modes is a block-diagonal matrix, containing 
a non-zero block $\tilde{\boldsymbol{H}}^{\rm sub,i}$ for each subset of localized modes along its diagonal. 

Different strategies can be used for assigning normal modes to subset in the localization procedure.
The character of normal modes (i.e., bending, stretching etc.) can be used to decide which modes 
should be localized together. Furthermore, the frequencies of normal modes can be taken into account, 
and thus modes lying in a given region of the spectrum are grouped and localized together. 
Previously \cite{jacob_localizing_2009, jacob_analysis_2009, jacob_understanding_2009,liegeois_analysis_2010,
weymuth_local-mode_2010}, such criteria have been used  to guide a normal mode assignment. We employ 
such criteria here for localizing the normal modes in subsets containing similar vibrations.

\section{\mbox{L-VSCF} and \mbox{L-VCI}}
\label{sec:lvscf_lvci}

We are aiming to solve the vibrational Schr\"odinger equation with the Watson Hamiltonian for a non-rotating molecule 
\cite{watson_simplification_1968}, 
\begin{align}
\label{eq:watson}
\hat{H} = \frac{1}{2}\sum_{\alpha\beta}\hat{\pi}_\alpha \mu_{\alpha\beta} \hat{\pi}_\beta - \frac{1}{8} \sum_\alpha 
\mu_{\alpha\alpha} - \frac{1}{2} \sum_i^M \frac{\partial^2}{\partial q_i^2} + V({\boldsymbol{q}}),
\end{align}
where the first term corresponds to the vibrational angular momentum (VAM), and the second is so-called Watson correction term,
both together are often referred as VAM terms.
Here, $\boldsymbol{\mu}$ is the inverse moment of the inertia tensor, and $\hat{\boldsymbol{\pi}}$ is the momentum operator. 
Since the exact evaluation of the VAM terms is cumbersome several approximations are commonly introduced, such as
a hierarchical expansion of the $\boldsymbol{\mu}$ tensor, or a selective inclusion of the terms in the VCI calculations
in combination with prescreening techniques \cite{carter_extensions_1998,neff_convergence_2011}.
As the VAM terms depend on the inverse of the inertia tensor, their contribution will decrease with increasing 
size of the molecule. Thus, for simplicity and efficiency, they are often omitted entirely, as we proceed here,
\begin{align}
\hat{H}= -\frac{1}{2} \sum_i^M \frac{\partial^2}{\partial {q}_i^2} + V({\boldsymbol{q}}),
\label{eq:vibham}
\end{align}
where $V({\boldsymbol{q}})$ is expressed as a truncated $n$-mode expansion in either the normal-mode coordinates
$\boldsymbol{q}$ or localized-mode coordinates $\tilde{\boldsymbol{q}}$.

For the approximate solution of the vibrational Schr\"odinger equation, the VSCF and VCI methods in normal-mode coordinates
are well established \cite{christiansen_selected_2012, roy_vibrational_2013}. Recently, we have introduced the analogous
localized-mode variants \mbox{L-VSCF} and \mbox{L-VCI} \cite{panek_efficient_2014}. Here, we briefly recall the essential steps.

The vibrational wavefunction in \mbox{L-VSCF} is given by the product ansatz,
\begin{align}
\label{eq:vscf2}
\Psi_n(\tilde{\boldsymbol{q}}) \approx \psi_{\boldsymbol{n}}(\tilde{q}_1,\ldots,\tilde{q}_M) = \prod_i^M \phi_i^{n_i}(\tilde{q}_i).
\end{align}
Here, $\phi_i^{n_i}(\tilde{q}_i)$ is a so-called modal for the $i$-th localized mode $\tilde{q}_i$ and $n_i$ is the 
vibrational quantum number for this modal.
Applying the variational principle to the nuclear Schr\"odinger equation, a set of one-mode equations is obtained,
\begin{align}
\label{eq:vscf4}
\hat{h}_i^{\boldsymbol{n}}(\tilde{q}_i) \, \phi_i^{n_i}(\tilde{q}_i) = \epsilon_i^{n_i} \, \phi_i^{n_i}(\tilde{q}_i),
\end{align}
with the effective Hamiltonian, 
\begin{align}
\label{eq:vscf5}
\hat{h}_i^{\boldsymbol{n}}(\tilde{q}_i) = -\frac{1}{2}\frac{\partial^2}{\partial\tilde{q}_i^2} + V_i^{(1)}(\tilde{q}_i)  +V_i^{\boldsymbol{n}}(\tilde{q}_i),
\end{align}
containing an effective mean-field potential,
\begin{align}
\label{eq:vscf6}
V_i^{\boldsymbol{n}}(\tilde{q}_i) 
   =& \ \sum_{j}^M \left\langle \phi_j^{n_j} \left| V_{ij}^{(2)} \right| \phi_j^{n_j} \right\rangle 
   + \sum_{j<k}^M \left\langle \phi_j^{n_j}\phi_k^{n_k} \left| V_{ijk}^{(3)} \right| \phi_j^{n_j}\phi_k^{n_k} \right
   \rangle \nonumber \\
   &+ \sum_{j<k<l}^M \left\langle \phi_j^{n_j}\phi_k^{n_k} \phi_l^{n_l} \left| V_{ijkl}^{(4)} \right| 
	\phi_j^{n_j}\phi_k^{n_k} \phi_l^{n_l} \right
   \rangle + \dotsb.
\end{align}
The resulting eigenvalue equations are then solved in a self-consistent manner to obtain the modal energies
$\epsilon_i^{n_i}$ as well as the optimized modals $\phi_i^{n_i}(\tilde{q}_i)$ themselves.

The correlation energy is then treated using (L-)VCI.\cite{bowman_application_1979} In this method, a total 
wavefunction for a considered vibrational state is built as a linear combination of $N_\text{states}$ \mbox{L-VCI} 
basis states, 
\begin{align}
\label{eq:vciwfn}
\Psi_i^{\rm \mbox{L-VCI}}(\tilde{\boldsymbol{q}}) =  \sum \limits_I^{N_\text{states}} c_I^{(i)} \psi^0_{\boldsymbol{n}_I}(\tilde{\boldsymbol{q}})
\end{align}
The functions $\psi_{\boldsymbol{n}_I}^0$ are products of the ground-state optimized \mbox{L-VSCF} modals 
$\phi_i^{n_i,0}$. Applying the variational principle with the ansatz of Eq.~\eqref{eq:vciwfn} leads to 
a CI eigenvalue equation with the CI-matrix,
\begin{align}
 \label{eq:vci1}
   \Bra{\psi^0_{\boldsymbol{n_I}}}\hat{H} \Ket{\psi^0_{\boldsymbol{n_J}}} 
      =&  \sum_i^M \left\langle \phi_i^{n_I,0} \left| -\frac{1}{2}\frac{\partial^2}{\partial \tilde{q}_i^2} 
        + V_i^{(1)} \right| \phi_i^{n_J,0} \right\rangle \prod_{j\neq i}^M \delta_{n_j^I n_j^J} \nonumber \\
        &+  \sum_{i<j}^M \left\langle \phi_i^{n_I,0}\phi_j^{n_I,0} \left| V_{ij}^{(2)} \right| \phi_i^{n_J,0}\phi_j^{n_J,0} 
        \right\rangle \prod_{k\neq i,j}^M \delta_{n_k^I n_k^J} \nonumber \\
        &+  \sum_{i<j<k}^M \left\langle \phi_i^{n_I,0}\phi_j^{n_I,0} \phi_k^{n_I,0} \left| V_{ijk}^{(3)} \right| 
        \phi_i^{n_J,0}\phi_j^{n_J,0} \phi_k^{n_J,0} \right\rangle \prod_{l\neq i,j,k}^M \delta_{n_l^I n_l^J} \nonumber \\
       &+ \sum_{i<j<k<l}^M \left\langle \phi_i^{n_I,0}\phi_j^{n_I,0} \phi_k^{n_I,0}\phi_l^{n_I,0} \left| V_{ijkl}^{(4)} \right| 
        \phi_i^{n_J,0}\phi_j^{n_J,0} \phi_k^{n_J,0}\phi_l^{n_J,0} \right\rangle \prod_{m\neq i,j,k,l}^M \delta_{n_m^I n_m^J} \nonumber \\
        &+ \dotsb
\end{align}
Its eigenvalues and eigenvectors are the requested \mbox{L-VCI} energies and \mbox{L-VCI} wavefunction coefficients, respectively.
The CI expansion in Eq.~\eqref{eq:vciwfn} is limited to a given order of excitation. 
The inclusion of higher excitations leads towards the limit of the full VCI (FVCI), where all possible
excitations are considered, but simultaneously increases the computational effort to construct and 
diagonalize the CI-matrix. 

There are several ways of constructing the excitation space in VCI calculations.
Christiansen proposed the most straightforward definition, denoted as VCI[$n$], where for each of the $M$ modals, 
excitations up to the $n$-th excited state are allowed \cite{christiansen_vibrational_2004}. 
Such a definition yields a rather large excitation space, and thus further limitations
are introduced. Rauhut defines the number of simultaneously excited modals by analogy to electronic structure calculations
giving VCI-S, VCI-SD, VCI-SDT, \ldots, where up to one, two, and three modals are excited, respectively 
\cite{rauhut_configuration_2007}. Additionally, two parameters $n_{\rm max}^1$ and $n_{\rm max}^\Sigma$ can be defined, 
limiting the maximal excitation per modal and the total sum of excitation quanta, respectively. Here, we equate the 
three parameters introduced by Rauhut. Thus, in our nomenclature VCI-SDTQ corresponds to a VCI space where
up to four mode are excited simultaneously with up to four excitation quanta, $n_{\rm max}^1=n_{\rm max}^\Sigma=4$.
Yagi \textit{et al.} reported that such a definition, denoted by them as VCI-($k$), is rational for VCI in the basis of  
optimized modals \cite{yagi_optimized_2012}.

In our calculations, the excitation space is constructed from a single VSCF reference, which was optimized for the vibrational
ground state. This approach is referred to as ground-state VCI (gs-VCI). Another possibility is to use a different VSCF
reference for each state of interest, which was optimized in this particular state, denoted state-specific VCI (ss-VCI). 
Such calculations should usually provide a faster convergence with respect to the excitation space \cite{christiansen_beyond_2005}.
However, it requires more computational effort, and yields non-orthogonal VCI wave functions.

\section{Computational details}
\label{sec:comp}

For ethene and furan we have used the normal modes, harmonic frequencies and anharmonic potential energy surfaces 
in terms of normal-mode coordinates available in the on-line Database of Potential Energy Surfaces maintained by Rauhut 
and co-workers \cite{online_pes_database,rauhut_efficient_2004,pfeiffer_anharmonic_2013}.
These harmonic and anharmonic potential energy surfaces were constructed by means of highly accurate explicitly
correlated CCSD(T)-F12x methods \cite{pfeiffer_anharmonic_2013}. Namely, the CCSD(T)-F12b method with a cc-pVTZ-F12 basis set
was used to obtain the Hessian and the anharmonic one-mode potentials, whereas CCSD(T)-F12a with a cc-pVDZ-F12 basis set was used
for the anharmonic two-mode and higher-order potentials \cite{adler_simple_2007,knizia_simplified_2009,peterson_systematically_2008}. 
For constructing anharmonic potential energy surfaces in terms of localized modes, we have localized the CCSD(T)-F12b/cc-pVTZ-F12
normal modes using our LocVib package \cite{jacob_localizing_2009,movipac}. Subsequently, we used the same methods as Rauhut
and co-workers to calculate the one-mode as well as (for ethene) the two-mode and three-mode potentials in terms of localized-mode 
coordinates with the \textsc{Molpro} 2012.1 program package \cite{molpro}.

To investigate the VCI convergence with respect to the $n$-mode expansion of the PES for the four highest modes of ethene
(see Section~\ref{sec:ethene_nmode_ch-str}), both for normal and localized modes, up to four-mode anharmonic potentials 
were obtained using density-functional theory (DFT) with the \textsc{Turbomole} 6.3.1 program package \cite{turbomole-1,turbomole-2}. 
The BP86 exchange--correlation functional \cite{B88,Perdew86} with Ahlrichs' def2-TZVP basis sets \cite{weigend_balanced_2005} in 
combination with the resolution-of-identity (RI) approximation and suitable auxiliary basis set was used \cite{turbomole-jbasen-1,
weigend_accurate_2006}.

Additionally, for furan we have constructed so-called hybrid potential energy surfaces, where the lower-order surfaces have been 
calculated with higher accuracy, whereas the higher-order surfaces have been calculated with lower accuracy \cite{rauhut_efficient_2004,
puzzarini_accurate_2010}.  We use a hybrid PES in terms of both normal-mode and localized-mode coordinates. Here, CCSD(T)/F12b
 is used to calculate the one-mode potentials, while the two-mode potentials are obtained with DFT (BP/def2-TZVP) as described above. 
Further details are discussed in Section~\ref{sec:furan_potentials}.

(L-)VSCF/(L-)VCI calcualtions were carried out with our Python code \textsc{Vibrations} \cite{panek_efficient_2014}.
All calculations were performed on equally spaced 16-point grids (see Ref.~\cite{panek_efficient_2014} for further details). 
\textsc{Vibrations} allows to perform VSCF and VCI calculations with user-defined vibrational coordinates.
The calculations can be carried out for all vibrational modes, as well as for a chosen subset of those. For constructing potential energy surfaces,
our code uses PyADF \cite{PyADF} as an interface to various quantum-chemistry packages. VCI calculations are performed in the basis of ground-state
optimized VSCF wave function, within a freely defined excitation space. Large Hamiltonian matrices can be efficiently constructed
using parallel techniques, distributing the work over many cores, whereas the diagonalization can be effectively performed in an iterative manner,
utilizing tools available in the \textsc{SciPy} \cite{scipy} and \textsc{NumPy} \cite{NumPy} packages.

\section{Test case: ethene}
\label{sec:ethene} 

To explore the advantages of \mbox{L-VSCF}/\mbox{L-VCI} over a conventional VSCF/VCI treatment, we consider the
ethene molecule as a first test case. Ethene has been extensively used in benchmarks of different methods for
calculating anharmonic vibrational frequencies, and therefore rich reference data are available \cite{martin_geometry_1996,
neugebauer_fundamental_2003,barone_anharmonic_2005,christiansen_beyond_2005,pfeiffer_anharmonic_2013}. 

\subsection{Normal and localized modes}
\label{sec:ethene_harmonic}

\begin{figure}
\caption{(a) Normal and (b) localized modes for ethene (CCSD(T)-F12b/cc-pVTZ-F12). 
The localization was carried out in the three subsets also indicated here.}
\label{fig:eth_modes}
\begin{center}
\includegraphics[width=0.8\textwidth]{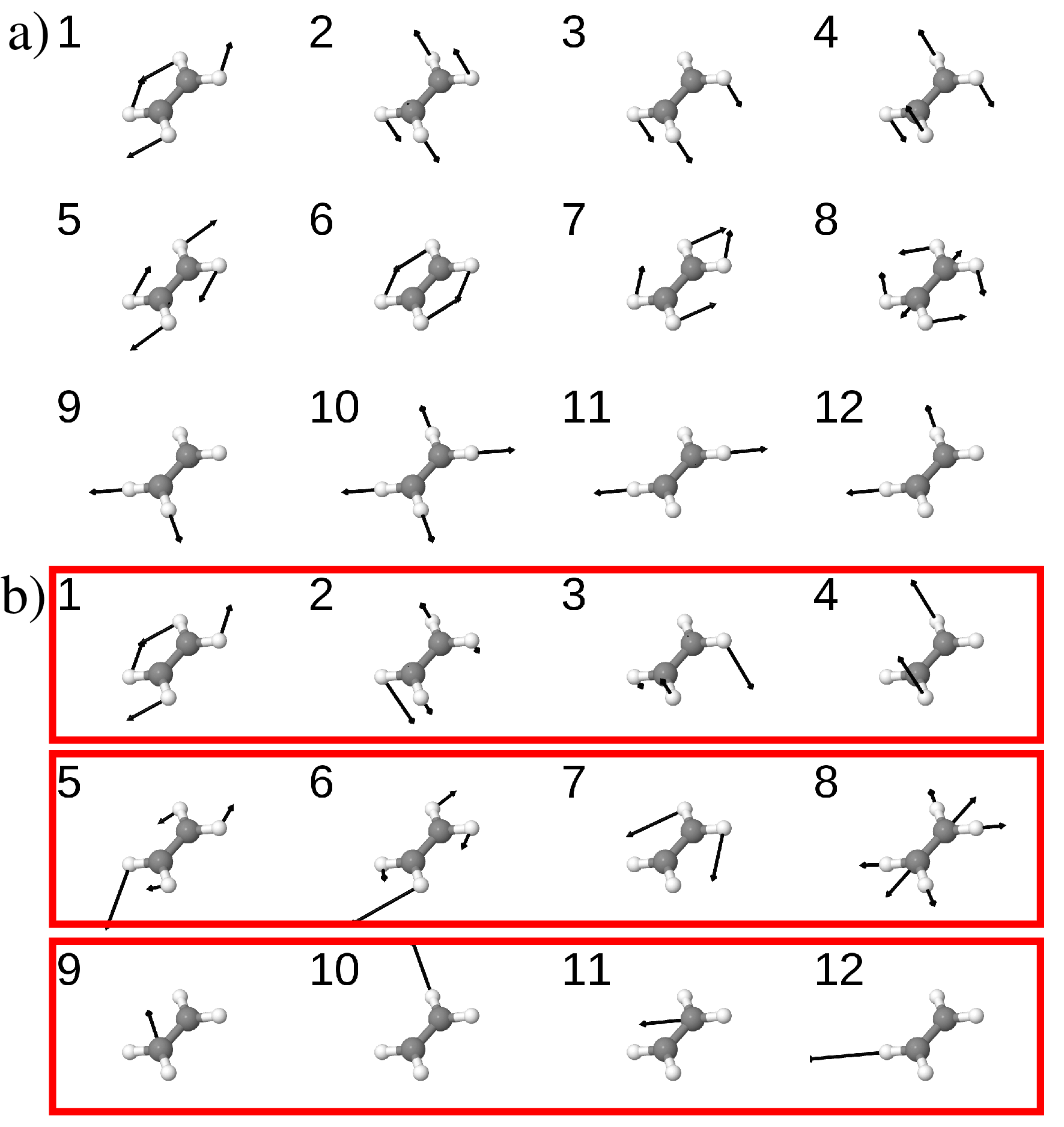}
\end{center}
\end{figure}

Graphical representations of the normal modes of the ethene molecule are shown in Fig.~\ref{fig:eth_modes}a,
and the corresponding harmonic vibrational frequencies are listed in Table~\ref{tab:eth_harmonic}. 
The first eight modes correspond to in-plane and out-of-plane bending vibrations, while the final four are
C--H stretching modes. For the localization, we have assigned the normal modes to three subsets, each consisting 
of four modes corresponding to characteristic groups of vibrations. This assignment as well as the resulting
localized modes are presented in Fig.~\ref{fig:eth_modes}b. Visually, the main character of the first eight modes 
is essentially unchanged upon localization, but in many cases there is a more distinct contribution of one single
vibration. On the other hand, the collective C--H stretching vibrations are decomposed into four distinct single-bond 
C--H stretching vibrations by the localization.

The fictitious vibrational frequencies of the localized modes (see Table~\ref{tab:eth_harmonic}) are on average 39~\rcm\ off 
from the harmonic frequencies. However, a \mbox{L-VCI}-S calculation using the harmonic PES expressed 
in localized-mode coordinates recovers the original harmonic frequencies within at most 2~\rcm. 

\begin{table}
\caption{Normal-mode ($\nu$) and localized-mode ($\tilde{\nu}$) vibrational frequencies for ethene (CCSD(T)-F12b/cc-pVTZ-F12).
The results of \mbox{L-VCI}-S using the harmonic PES expressed in localized-mode coordinates are also included,
along with the mean average deviation (MAD) and their maximum absolute deviation (MAX) from the normal-mode harmonic frequencies. 
All frequencies are given in~\rcm.}
\label{tab:eth_harmonic}
\begin{center}
\begin{spacing}{1.3}
\begin{tabular}{r|r|rrrr}
\hline\hline
\multicolumn{1}{c|}{Mode} & \multicolumn{1}{c|}{Normal} & \multicolumn{4}{c}{Localized Modes} \\ \hline
 & $\nu$ & \multicolumn{2}{c}{$\tilde{\nu}$} & \multicolumn{2}{c}{\mbox{L-VCI}-S}  \\ \hline
1 & 825 & 825 & 0 & 825 & 0 \\
2 & 949 & 977 & 28 & 949 & -1 \\
3 & 963 & 977 & 14 & 963 & 0 \\
4 & 1050 & 1008 & -42 & 1051 & 1 \\ \cline{3-6}
5 & 1248 & 1354 & 106 & 1246 & -2 \\
6 & 1368 & 1354 & -15 & 1369 & 1 \\
7 & 1477 & 1462 & -15 & 1476 & -2 \\
8 & 1671 & 1595 & -76 & 1673 & 2 \\ \cline{3-6}
9 & 3140 & 3191 & 51 & 3140 & 0 \\
10 & 3155 & 3191 & 36 & 3155 & 0 \\
11 & 3222 & 3191 & -31 & 3222 & 0 \\
12 & 3248 & 3191 & -57 & 3248 & 0 \\ \hline
   \multicolumn{2}{c}{MAD} &  & 39 &  & 1 \\
   \multicolumn{2}{c}{MAX} &  & 106 &  & 2 \\ 
 \hline\hline
\end{tabular}
\end{spacing}
\end{center}
\end{table}

Before turning to the full anharmonic calculations for ethene, we investigate how the choice of normal-mode coordinates
or localized-mode coordinates affects the coupling between the modes via the two-mode potentials. To this end, the magnitude 
of the coupling between modes $i$ and $j$ is calculated as the absolute value of the expectation value of the two-mode potential 
operator for modes $i$ and $j$ with the ground-state optimized VSCF wave function [cf.~Eq.~\eqref{eq:vscf6}],
\begin{align}
C(i,j) = \left| \left\langle \phi_i^{0,0}\phi_j^{0,0} \left| V_{ij}^{(2)} \right| \phi_i^{0,0}\phi_j^{0,0} 
        \right\rangle \right|,
\end{align}
where $\phi_i^{0,0}$ is the ground-state optimized VSCF modal for the $i$-th mode in its ground-state.

\begin{figure}
\caption{Strengths of the two-mode couplings, $C(i,j)$, for the normal and localized modes (central figure) and total coupling of the $i$-th normal 
and localized mode with all other modes (bar plots), $C(i)$, for ethene. The lower right part refers to normal modes, whereas the upper left 
part refers to localized modes. See text for further details.}
\label{fig:ethene_nmvslm}
\begin{center}
\includegraphics[width=\textwidth]{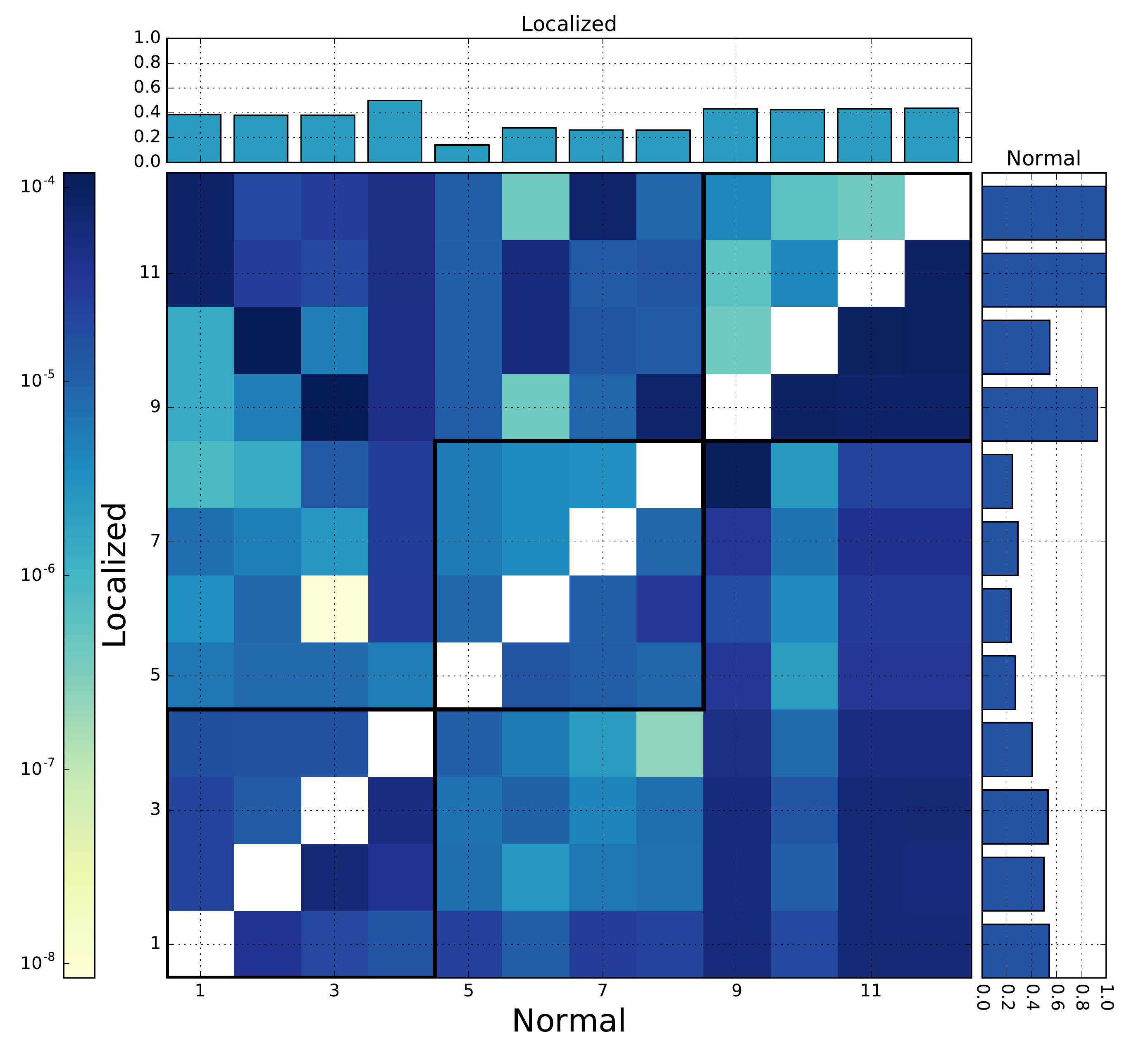}
\end{center}
\end{figure}

These couplings are visualized in Fig.~\ref{fig:ethene_nmvslm}. Here, the upper triangle of the matrix corresponds to the localized 
modes, whereas the lower triangle corresponds to the normal modes. The black boxes indicate modes that were localized together. 
Noticeably, the coupling between the localized modes within subsets is smaller than for the corresponding normal modes, which is 
especially pronounced for the C--H stretching modes (modes 9--12).  
The bar charts present the total coupling of the $i$-th normal or localized mode with other modes, calculated as $C(i)=\sum_{j\neq i}C(i,j)$. 
For these total couplings of C--H stretching vibrations, one can notice that the localized modes are not only less coupled within their subsets, 
but that they are also less coupled with all other modes. Overall, the localized modes show weaker couplings, with some stronger individual 
spots, whereas the normal modes have rather uniformly strong couplings. Thus, already for a molecule as small as ethene, switching from 
normal modes to localized modes can reduce the two-mode couplings. These benefits can be expected to become larger as the size of the 
considered molecule increases \cite{panek_efficient_2014}, especially for the lower-frequency modes.

\subsection{Convergence of the $n$-mode expansion: C--H stretching modes}
\label{sec:ethene_nmode_ch-str}

To analyze the convergence of (L-)VCI calculations with respect to the order of the \mbox{$n$-mode} expansion,
we initially consider only one subset of normal modes, namely the four C--H stretching vibrations. In this subset, 
for both normal and localized modes, all potentials up to the fourth order were calculated by means of density-functional 
theory (DFT), using grids with 16 points along each mode. The construction of this approximated PES required
83~520 single-point calculations in total (64, 1~536, 16~384, and 65~536, for $V^{(1)}$, $V^{(2)}$, $V^{(3)}$, and $V^{(4)}$,
respectively). These potentials were employed in VCI and \mbox{L-VCI} calculations, and up to quintuple excitations 
were considered [(L-)VCI-SDTQ5]. The results are shown in Table~\ref{tab:eth_anh_convergence}. 

If the $n$-mode expansion is performed in terms of normal-mode coordinates, it is clear that up to four-mode potentials 
need to be included. Neglecting the four-mode potentials and including only the contributions up to $V^{(3)}$ leads to
deviations of up to 20~\rcm\ in the resulting vibrational frequencies. This changes if the PES expanded in terms of 
localized-mode coordinates. In this case, already \mbox{L-VCI} with only one-mode and two-mode potentials yields converged 
results, with deviations of at most 1~\rcm\ from the normal-modes reference. A significant difference can be observed
in the VCI expansions for both cases. For the normal modes, each of the converged C--H transitions consists to at least 95\%
of the respective singly-excited VSCF state. In the localized-mode basis, these transitions are combinations of the four
singly excited L-VSCF states with roughly equal weights. 
Furthermore, it is remarkable that already \mbox{L-VCI-S} (i.e., including only single excitations) gave vibrational frequencies 
being within~4~\rcm\ agreement with the values presented in the table. Thus, when using localized-mode coordinates instead of 
normal-mode coordinates, only 1600 single-point calculations (instead of 83~520 single-point calculations) and a significantly 
smaller excitation space (VCI-S instead of VCI-SDTQ5) are necessary to reach the same accuracy. Note that these one-mode
potentials in localized-mode coordinates are transformed into up to four-mode potentials when going to normal modes because 
each localized modes is a linear combination of four normal modes.

\begin{table}
\caption{Convergence of the (L-)VCI-SDTQ5 fundamental vibrational frequencies for the C--H stretching vibrations 
in ethene (DFT/BP/def2-TZVP) with respect to the order of the $n$-mode expansion. $V^{(\lbrace n\rbrace)}$ denotes 
the orders of the potentials included in the respective calculations. Only the subset of the C--H stretching vibrations is 
considered here and all couplings to other modes are neglected. For the frequencies obtained with up to two- and
three-mode potentials, the deviations, mean absolute deviations (MAD), and maximal absolute deviations (MAX) with 
respect to the results including up to four-mode potentials are also included. All frequencies are given in~\rcm.}
\label{tab:eth_anh_convergence}
\begin{center}
\begin{spacing}{1.3}
\begin{tabular}{l|rr|rr|r|rr|rr|r}
\hline\hline
&\multicolumn{5}{c|}{Normal Modes} & \multicolumn{5}{c}{Localized Modes} \\ \hline
Mode &\multicolumn{2}{c|}{$V^{(1,2)}$} & \multicolumn{2}{c|}{$V^{(1,2,3)}$}  & \multicolumn{1}{c|}{$V^{(1,2,3,4)}$}  & \multicolumn{2}{c|}{$V^{(1,2)}$} & 
\multicolumn{2}{c|}{$V^{(1,2,3)}$}  & \multicolumn{1}{c}{$V^{(1,2,3,4)}$} \\ \hline
9 &2969 & 19 & 2930 & -20 & 2950 & 2949 & 0 & 2949 & 0 & 2949 \\
10 &3010 & 49 & 2954 &  -7 & 2961 & 2960 & 1 & 2961 & 0 & 2961 \\
11 &3073 & 62 & 2991 & -20 & 3011 & 3010 & 0 & 3010 & 0 & 3010 \\
12 &3104 & 63 & 3020 & -21 & 3041 & 3040 & 0 & 3040 & 0 & 3040 \\ \hline
MAD  && 48 &      &  17 &      &      & 0 &      & 0 & \\
MAX  && 63 &      &  21 &      &      & 1 &      & 0 & \\
\hline\hline
\end{tabular}
\end{spacing}
\end{center}
\end{table}

\subsection{Convergence of the VCI expansion}
\label{sec:ethene_vci}

Next, we consider all modes and investigate the convergence of the fundamental frequencies with respect to the excitation space 
used in the (L-)VCI Hamiltonian. Here, the PES was approximated by the $n$-mode expansion in terms of normal-mode
coordinates or localized-mode coordinates, truncated after two-mode or three-mode potentials. 
For both the expansion in normal-mode coordinates and in localized-mode coordinates, the fundamental frequencies obtained at
each VCI excitation level as well as their deviations from the reference values obtained at the (L-)VCI-SDTQ56 level are listed
in Tables~\ref{tab:eth_anh_1}~and~\ref{tab:eth_anh_2}, respectively.

\begin{table}
\caption{Convergence of the VCI-SDTQ5 fundamental vibrational frequencies of ethene (CCSD(T)-F12x) 
with respect to the order of the VCI expansion. In each column, the vibrational frequencies are given together with their deviations from 
the \mbox{VCI-SDTQ56} value as well as the mean average deviation (MAD) and their maximum absolute deviation (MAX). 
The calculations have been performed for different truncations of the $n$-mode expansion in terms of normal-mode coordinates. 
$V^{(\lbrace n\rbrace)}$ denotes the orders of the potentials included in the respective calculations.
All frequencies are given in~\rcm.}
\label{tab:eth_anh_1}
\begin{center}
\begin{spacing}{1.2}
\begin{tabular}{r|rr|rr|rr|rr|rr|r}\hline\hline
\multicolumn{12}{c}{Normal Modes, $V^{(1,2)}$} \\ \hline 
\multicolumn{1}{c|}{Mode} & \multicolumn{2}{c|}{VCI-S} & \multicolumn{2}{c|}{-SD} & \multicolumn{2}{c|}{-SDT} & \multicolumn{2}{c|}{-SDTQ} & \multicolumn{2}{c|}{-SDTQ5} & \multicolumn{1}{c}{-SDTQ56} \\ \hline
1 & 847 & 16 & 839 & 8 & 867 & 37 & 834 & 4 & 830 & 0 & 830 \\
2 & 951 & 11 & 946 & 6 & 974 & 34 & 943 & 3 & 940 & 0 & 940 \\
3 & 966 & 13 & 959 & 7 & 988 & 35 & 956 & 3 & 952 & 0 & 952 \\
4 & 1037 & 8 & 1033 & 4 & 1062 & 33 & 1031 & 3 & 1029 & 0 & 1029 \\
5 & 1235 & 3 & 1233 & 2 & 1262 & 31 & 1233 & 2 & 1231 & 0 & 1231 \\
6 & 1350 & 8 & 1345 & 3 & 1374 & 32 & 1344 & 2 & 1342 & 0 & 1342 \\
7 & 1450 & 3 & 1449 & 2 & 1478 & 31 & 1449 & 2 & 1447 & 0 & 1447 \\
8 & 1642 & 15 & 1633 & 6 & 1661 & 34 & 1632 & 5 & 1627 & 1 & 1627 \\
9 & 3058 & 54 & 3021 & 17 & 3049 & 45 & 3010 & 6 & 3005 & 0 & 3004 \\
10 & 3059 & 52 & 3033 & 26 & 3061 & 53 & 3015 & 7 & 3008 & 0 & 3008 \\
11 & 3131 & 59 & 3087 & 16 & 3114 & 43 & 3074 & 3 & 3071 & 0 & 3071 \\
12 & 3157 & 58 & 3114 & 16 & 3141 & 43 & 3101 & 3 & 3099 & 0 & 3098 \\
\hline
\multicolumn{2}{c}{MAD} & 25 &  & 9 &  & 38 &  & 4 &  & 0\\
\multicolumn{2}{c}{MAX} & 59 &  & 26 &  & 53 &  & 7 &  & 1 \\ \hline
\multicolumn{12}{c}{Normal Modes, $V^{(1,2,3)}$} \\ \hline 
1 & 850 & 28 & 832 & 11 & 895 & 73 & 825 & 4 & 821 & -1 & 821 \\
2 & 955 & 26 & 938 & 9 & 1001 & 72 & 933 & 3 & 929 & -1 & 929 \\
3 & 970 & 29 & 952 & 10 & 1014 & 73 & 945 & 4 & 941 & -1 & 942 \\
4 & 1041 & 23 & 1026 & 8 & 1089 & 71 & 1021 & 3 & 1017 & -1 & 1018 \\
5 & 1237 & 14 & 1227 & 4 & 1291 & 67 & 1225 & 2 & 1222 & -1 & 1223 \\
6 & 1352 & 11 & 1344 & 3 & 1407 & 66 & 1344 & 2 & 1340 & -1 & 1341 \\
7 & 1454 & 17 & 1442 & 5 & 1505 & 68 & 1440 & 3 & 1437 & -1 & 1437 \\
8 & 1644 & 21 & 1630 & 8 & 1692 & 70 & 1629 & 7 & 1622 & 0 & 1622 \\
9 & 3060 & 106 & 3004 & 51 & 3059 & 106 & 2975 & 22 & 2955 & 2 & 2953 \\
10 & 3067 & 71 & 3043 & 46 & 3096 & 99 & 3016 & 19 & 3001 & 4 & 2997 \\
11 & 3140 & 93 & 3106 & 59 & 3160 & 113 & 3070 & 23 & 3050 & 3 & 3047 \\
12 & 3166 & 100 & 3126 & 60 & 3180 & 114 & 3085 & 18 & 3069 & 2 & 3066 \\
\hline
\multicolumn{2}{c}{MAD} & 45 &  & 23 &  & 83 &  & 9 &  & 1 \\
\multicolumn{2}{c}{MAX }& 106 &  & 60 &  & 114 &  & 23 &  & 4 \\
\hline\hline
\end{tabular}
\end{spacing}
\end{center}
\end{table}

\begin{table}
\caption{Convergence of the L-VCI-SDTQ5 fundamental vibrational frequencies of ethene (CCSD(T)-F12x) 
with respect to the order of the VCI expansion. In each column, the vibrational frequencies are given together with their deviations from 
the \mbox{L-VCI-SDTQ56} value as well as the mean average deviation (MAD) and their maximum absolute deviation (MAX). 
The calculations have been performed for different truncations of the $n$-mode expansion in terms of localized-mode coordinates. 
$V^{(\lbrace n\rbrace)}$ denotes the orders of the potentials included in the respective calculations.
All frequencies are given in~\rcm.}
\label{tab:eth_anh_2}
\begin{center}
\begin{spacing}{1.2}
\begin{tabular}{r|rr|rr|rr|rr|rr|r} \hline\hline
\multicolumn{1}{c|}{Mode} & \multicolumn{2}{c|}{\mbox{L-VCI}-S} & \multicolumn{2}{c|}{-SD} & \multicolumn{2}{c|}{-SDT} & \multicolumn{2}{c|}{-SDTQ} & \multicolumn{2}{c|}{-SDTQ5} & \multicolumn{1}{c}{-SDTQ56} \\ \hline
\multicolumn{12}{c}{Localized Modes, $V^{(1,2)}$ } \\ \hline
1 & 839 & 21 & 843 & 24 & 845 & 26 & 826 & 8 & 820 & 2 & 818 \\
2 & 938 & 25 & 940 & 27 & 940 & 28 & 922 & 9 & 915 & 2 & 913 \\
3 & 967 & 23 & 969 & 25 & 971 & 26 & 953 & 8 & 946 & 2 & 944 \\
4 & 1039 & 24 & 1040 & 26 & 1039 & 25 & 1022 & 8 & 1016 & 2 & 1014 \\
5 & 1226 & 12 & 1236 & 22 & 1234 & 20 & 1220 & 6 & 1215 & 1 & 1214 \\
6 & 1342 & 14 & 1351 & 23 & 1347 & 19 & 1334 & 6 & 1329 & 1 & 1328 \\
7 & 1458 & 10 & 1467 & 19 & 1466 & 18 & 1453 & 5 & 1449 & 1 & 1448 \\
8 & 1653 & 25 & 1659 & 31 & 1653 & 25 & 1641 & 13 & 1633 & 5 & 1628 \\
9 & 2961 & -10 & 2985 & 14 & 2988 & 17 & 2973 & 2 & 2974 & 3 & 2971 \\
10 & 2976 & -40 & 3039 & 22 & 3040 & 24 & 3024 & 8 & 3017 & 0 & 3017 \\
11 & 3045 & -11 & 3072 & 17 & 3074 & 19 & 3062 & 6 & 3057 & 1 & 3055 \\
12 & 3072 & -9 & 3099 & 17 & 3101 & 19 & 3088 & 6 & 3082 & 1 & 3081 \\
\hline
\multicolumn{2}{c}{MAD} & 19 &  & 22 &  & 22 &  & 7 &  & 2\\
\multicolumn{2}{c}{MAX} & 40 &  & 31 &  & 28 &  & 13 &  & 5 \\ \hline
\multicolumn{12}{c}{Localized Modes, $V^{(1,2,3)}$ } \\ \hline
1 & 850 & 33 & 845 & 27 & 855 & 38 & 830 & 12 & 820 & 3 & 817 \\
2 & 951 & 31 & 948 & 27 & 958 & 38 & 933 & 12 & 923 & 3 & 920 \\
3 & 970 & 30 & 966 & 26 & 977 & 36 & 952 & 12 & 943 & 3 & 940 \\
4 & 1043 & 25 & 1041 & 23 & 1050 & 33 & 1028 & 10 & 1020 & 2 & 1017 \\
5 & 1240 & 20 & 1243 & 23 & 1251 & 31 & 1229 & 9 & 1222 & 2 & 1220 \\
6 & 1355 & 13 & 1361 & 18 & 1369 & 26 & 1351 & 8 & 1345 & 2 & 1343 \\
7 & 1454 & 20 & 1454 & 20 & 1464 & 30 & 1443 & 9 & 1436 & 2 & 1434 \\
8 & 1645 & 26 & 1645 & 26 & 1650 & 31 & 1631 & 12 & 1622 & 4 & 1619 \\
9 & 2974 & -2 & 2996 & 20 & 3011 & 35 & 2991 & 14 & 2980 & 4 & 2976 \\
10 & 2990 & -23 & 3033 & 20 & 3045 & 32 & 3027 & 15 & 3018 & 5 & 3012 \\
11 & 3039 & -37 & 3095 & 19 & 3107 & 31 & 3089 & 13 & 3080 & 4 & 3076 \\
12 & 3066 & -32 & 3116 & 18 & 3128 & 29 & 3109 & 11 & 3101 & 2 & 3098 \\
\hline
\multicolumn{2}{c}{MAD} & 24 &  & 22 &  & 32 &  & 11 &  & 3 \\
\multicolumn{2}{c}{MAX} & 37 &  & 27 &  & 38 &  & 15 &  & 5 \\
\hline\hline
\end{tabular}
\end{spacing}
\end{center}
\end{table}

Looking at the results in normal-mode coordinates (cf.~Table~\ref{tab:eth_anh_1}), the calculations with up to two-mode potentials
are converged within 1~\rcm\ when quintuple excitation are included. The initial VCI-S fundamentals are on average
25~\rcm\ off from the reference, while the mean deviation decreases to only 9~\rcm\ when including double excitations. 
However, when going to VCI-SDT there is a pronounced increase of the mean deviation to 38~\rcm.
In general, the convergence is faster for the low-frequency bending modes, whereas the C--H stretching modes converge 
significantly slower. A similar convergence pattern is observed for the calculations including up to three-mode potentials. 
Here, however, the deviations for each excitation level are roughly twice as high as for the PES including
only two-mode potentials. Moreover, the VCI-SDTQ5 C--H stretching frequencies seem to be not fully converged, and an
even larger excitation space might be needed to reach full convergence.

In localized-mode coordinates (cf.~Table~\ref{tab:eth_anh_2}), the fundamental frequencies generally converge more smoothly. 
Already with \mbox{L-VCI}-S, the mean average deviation is smaller than when using normal-mode coordinates. Both for the calculations 
including up to two-mode and up to three-mode potentials, the deviations upon inclusion of triple excitations is smaller than in the normal 
modes case. Moreover, when going from the PES including only two-mode potentials to the one including
also three-mode potentials, the increase of the mean average deviation is smaller than for the expansion of the potential 
energy surface in normal-mode coordinates. On the other hand, also in localized-mode coordinates, the \mbox{L-VCI}-SDTQ5 
frequencies appear to be not fully converged.

\subsection{Convergence of the $n$-mode expansion for all fundamentals}
\label{sec:ethene_nmode_all}

In Table~\ref{tab:eth_anh_exp}, we compare the best-estimate results for VCI and \mbox{L-VCI} with different truncations of
the $n$-mode expansion of the PES to the experimental reference values. Here, the VCI excitation space contains up to 
sextuple excitations both for normal and localized modes. For VCI, we use the PES approximated with up to the four-mode 
terms, whereas for L-VCI only up to the three-mode terms are included.

\begin{table}
\caption{(L-)VCI-SDTQ56 fundamental vibrational frequencies calculated for ethene with different truncations of 
the $n$-mode expansion of the PES [CCSD(T)-F12x]. $V^{(\lbrace n\rbrace)}$ denotes the orders of the potentials 
included in the respective calculations, $^{\rm 2h}$ refers to two-mode potentials obtained within the harmonic 
approximation. In each column, the frequencies are given together with 
their deviations from the experimental reference value as well as the mean average deviation (MAD) and their 
maximum absolute deviation (MAX).
. All frequencies given in~\rcm.}
\label{tab:eth_anh_exp}
\begin{center}
\begin{spacing}{1.3}
\hspace*{-1cm}
\begin{tabular}{r|rr|rr|rr|rr|rr|rr|rr|r}
\hline\hline 
\multicolumn{1}{c|}{Mode} & \multicolumn{8}{c|}{Normal Modes}  & \multicolumn{6}{c|}{Localized Modes}  &  Exp.$^{\rm a}$\\ \cline{2-15}
 & \multicolumn{2}{c|}{$V^{(1)}$} & \multicolumn{2}{c|}{$V^{(1,2)}$} & \multicolumn{2}{c|}{$V^{(1,2,3)}$} & \multicolumn{2}{c|}{$V^{(1,2,3,4)}$} & \multicolumn{2}{c|}{$V^{(1,2{\rm h})}$} & \multicolumn{2}{c|}{$V^{(1,2)}$} & \multicolumn{2}{c|}{$V^{(1,2,3)}$} & \\ \hline
1 & 863 & 37 & 830 & 5 & 821 & -5 & 819 & -7 & 863 & 37 & 818 & -8 & 817 & -9 & 826 \\
2 & 975 & 35 & 940 & 0 & 929 & -11 & 927 & -13 & 990 & 50 & 913 & -27 & 920 & -20 & 940 \\
3 & 990 & 41 & 952 & 4 & 942 & -7 & 939 & -10 & 999 & 50 & 944 & -4 & 940 & -9 & 949 \\
4 & 1062 & 36 & 1029 & 3 & 1018 & -8 & 1016 & -9 & 1083 & 57 & 1014 & -11 & 1017 & -8 & 1026 \\
5 & 1253 & 31 & 1231 & 9 & 1223 & 1 & 1222 & 0 & 1257 & 35 & 1214 & -8 & 1220 & -2 & 1222 \\
6 & 1370 & 27 & 1342 & -2 & 1341 & -2 & 1340 & -3 & 1368 & 25 & 1328 & -16 & 1343 & -1 & 1344 \\
7 & 1480 & 38 & 1447 & 5 & 1437 & -5 & 1437 & -6 & 1483 & 40 & 1448 & 6 & 1434 & -8 & 1443 \\
8 & 1668 & 42 & 1627 & 1 & 1622 & -3 & 1621 & -5 & 1661 & 35 & 1628 & 3 & 1619 & -7 & 1625 \\
9 & 3128 & 140 & 3004 & 16 & 2953 & -35 & 2984 & -4 & 3018 & 30 & 2971 & -18 & 2976 & -12 & 2989 \\
10 & 3171 & 149 & 3008 & -14 & 2997 & -25 & 3019 & -3 & 3033 & 11 & 3017 & -5 & 3012 & -10 & 3022 \\
11 & 3258 & 174 & 3071 & -12 & 3047 & -36 & 3071 & -13 & 3102 & 18 & 3055 & -28 & 3076 & -7 & 3083 \\
12 & 3282 & 177 & 3098 & -6 & 3066 & -39 & 3094 & -11 & 3128 & 23 & 3081 & -24 & 3098 & -7 & 3105 \\ \hline
 & \multicolumn{1}{c}{MAD} & 77 &  & 6 &  & 15 &  & 7 &  & 34 &  & 13 &  & 8 \\
 & \multicolumn{1}{c}{MAX} & 177 &  & 16 &  & 39 &  & 13 &  & 57 &  & 28 &  & 20 \\
\hline\hline
\multicolumn{16}{l}{$^{\rm a}$ -- experimental values as referenced in Ref.~\cite{martin_geometry_1996}} 
\end{tabular}
\end{spacing}
\end{center}
\end{table}

Normal-mode VCI fundamental frequencies with the PES approximated with at most two-mode potentials have an 
MAD of 6~\rcm. When going to a more accurate PES including also the three-mode terms, the mean average deviation
increases to 15~\rcm. Here, the C--H stretching modes contribute the most to this discrepancy. Inclusion of the four-mode 
terms in the $n$-mode expansion reduces the MAD to 7~\rcm, predominantly for the C--H stretching modes.

Local-mode \mbox{L-VCI} with one-mode and two-mode potentials included delivers fundamental transition which are on 
average 13~\rcm\ off from the reference. This is a deviation of similar magnitude as for the normal-modes VCI with up to 
three-mode potential included. However, the problematic C--H stretching modes energies are slightly better reproduced in
the L-VCI case. Inclusion of the three-mode potentials in the PES expansion reduces the MAD to 8~\rcm, which in turn is 
similar to the MAD of normal-modes VCI including up to four-mode potentials. This resembles the previously observed faster 
convergence with respect to the order of the $n$-mode expansion for the subset of C--H stretching modes 
(cf.~Table~\ref{tab:eth_anh_convergence}).

Despite the use of highly accurate potential energy surfaces with up to four or three-mode potentials included,
there is still some deviation with respect to the experimental reference values. This is most likely due to the use of 
the simplified vibrational Hamiltonian given in Eq.~\ref{eq:vibham} (see also Table~II in Ref.~\cite{pfeiffer_anharmonic_2013}). 
Additionally, a rather compact excitation space was used and it might be necessary to include even higher excitations 
in the VCI expansion.

Finally, we also explore a low-cost model that could serve a first approximation of anharmonic corrections. Since the Hessian 
in the localized modes basis is not diagonal, the arising off-diagonal elements can be used to calculate harmonic two-mode 
potentials \cite{panek_efficient_2014}. These can be combined with the anharmonic one-mode potentials. This approximate
PES is labelled $V^{(1,2{\rm h})}$ in Table~\ref{tab:eth_anh_exp}. Strong discrepancies can be observed for the bending modes, 
with deviations of up to 57~\rcm. However, the C--H stretching modes are reproduced quite well and the overall MAD is 34~\rcm. 
If the PES is expanded in normal modes, with the same computational effort one can calculate only the anharmonic one-mode 
potentials, $V^{(1)}$. Here, the deviation for the bending modes is at a similar level, but the C--H stretching modes are at least 
140~\rcm\ off from the experiment. Thus, using localized modes in such a low-cost model, we reduce the error by over 130~\rcm\
for the C--H stretching vibrations, while staying in a similar error range for the bending modes.

In summary, calculations performed for ethene with localized modes showed that these coordinates yield fundamental transition energies
that are equivalent to those obtained with normal modes. However, with local modes a smaller expansion of the PES
could be applied. In localized-mode coordinates, already a PES including up to three-mode potentials gives fundamental
vibrational frequencies that are as accurate as those obtained with a PES expanded in normal-mode coordinates including
up to four-mode potentials. Thus, the number of required single-point energy calculations can be reduced from 
33\,358\,528 in the case of normal-mode coordinates to only 918\,208 in the case of localized mode coordinates. 
Moreover, localized modes seem to be more suitable for devising low-cost models that can give a first estimate of anharmonic 
effects by including only one-mode potentials and harmonic two-mode potentials. Finally, the convergence of the VCI expansion 
is smoother when using localized-mode coordinates, and smaller deviations are observed at low excitation level, which can 
again be exploited for devising low-cost models.

\section{Furan}
\label{sec:furan}

Another test system for our method was furan. It is a small heterocyclic aromatic molecule, in which in addition to the C--H stretching
vibrations also ring vibrations are present. This molecule has extensively been used to benchmark methods for anharmonic vibrational
calculations \cite{simandiras_correlated_1988,burcl_vibrational_2003,barone_vibrational_2004,
danecek_comparison_2007,biczysko_harmonic_2010,pfeiffer_anharmonic_2013}.

\subsection{Normal and localized modes}
\label{sec:furan_harmonic}

\begin{figure}
\caption{(a) Normal and (b) localized modes for furan (CCSD(T)-F12b/cc-pVTZ-F12). 
The localization was carried out in the subsets also indicated here.}
\label{fig:furan_modes}
\begin{center}
\includegraphics[width=0.8\textwidth]{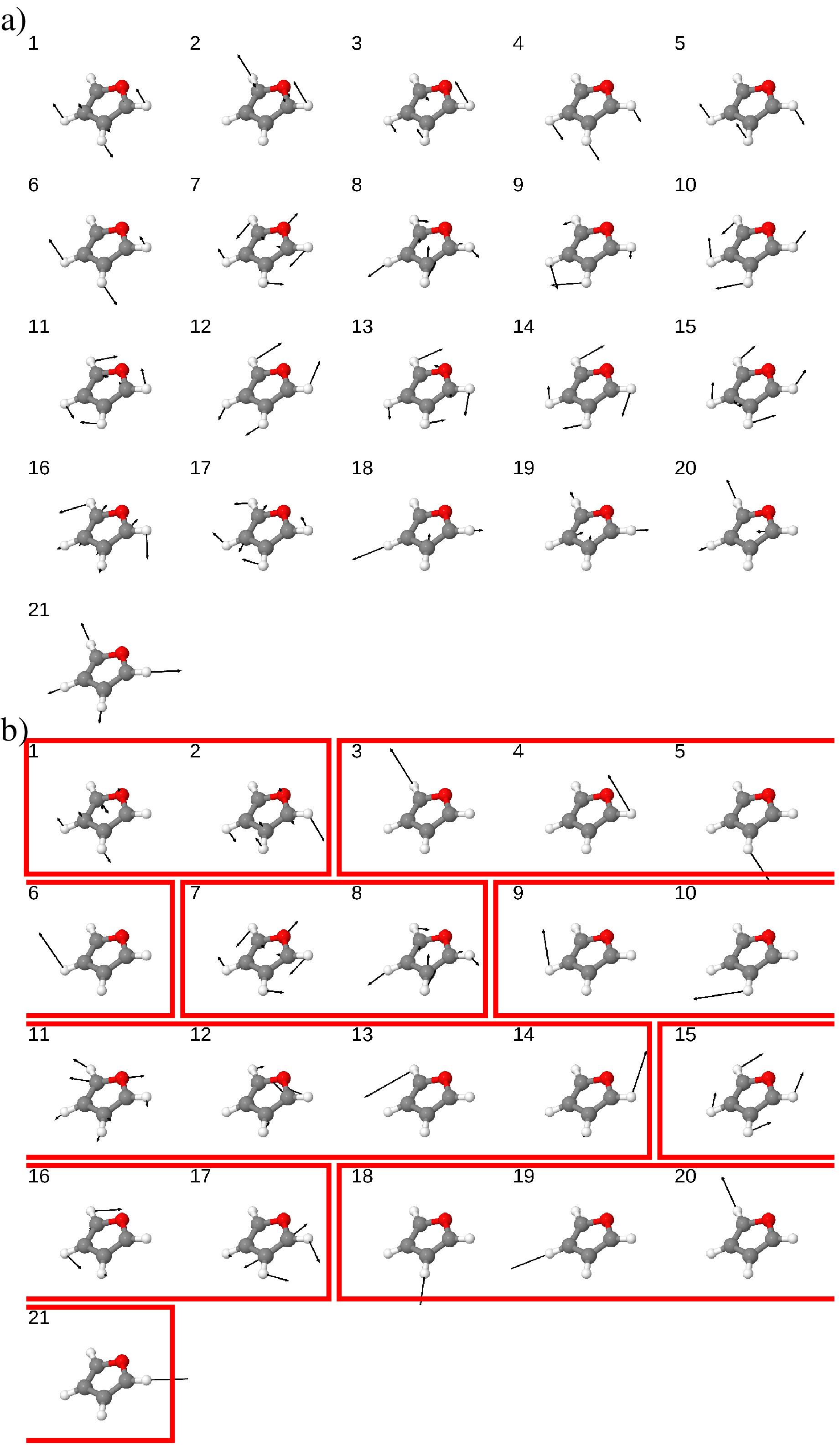}
\end{center}
\end{figure}

The CCSD(T)-F12b/cc-pVTZ normal modes for furan are shown in Fig.~\ref{fig:furan_modes}a and the corresponding
harmonic frequencies are listed in Table~\ref{tab:furan_harm}. For the localization, the normal modes have been assigned
to subsets according to their character and their harmonic frequencies. These subsets as well the the resulting localized
modes are shown in Fig.~\ref{fig:furan_modes}b, whereas the resulting fictitious harmonic frequencies of the localized
modes are listed in Table~\ref{tab:furan_harm}.

The normal modes are rather delocalized over the entire molecule and each normal mode involves many atoms. After the localization, 
some of them remain unchanged, or at least are still delocalized. However, in some cases a localization is observed, namely for the 
subsets of modes 3--6, modes 9--14 and modes 18--21. In the first of these subsets (modes 3--6), each mode corresponds to different 
out-of-plane bending of the C--H group. The second of these subsets (modes 9--14), are single C--H in-plane bending vibrations, except 
for modes~11 and~12 involving the oxygen atom. The last one of these subsets (modes 18--21) consists of single C--H stretching 
vibrations. \mbox{L-VCI}-S calculations using the harmonic PES expanded in localized-mode coordinates reproduce the initial 
normal-mode harmonic frequencies within on average 1~\rcm, with a maximal deviation of 6~\rcm\ for mode 14 (see Table~\ref{tab:furan_harm}).

\begin{table}
\caption{Normal-mode ($\nu$) and localized-mode ($\tilde{\nu}$) vibrational frequencies for furan (CCSD(T)-F12b/cc-pVTZ-F12).
The results of \mbox{L-VCI}-S using the harmonic PES expressed in localized-mode coordinates are also included,
along with the mean average deviation (MAD) and their maximum absolute deviation (MAX) from the normal-mode harmonic frequencies. 
All frequencies are given in~\rcm.}
\label{tab:furan_harm}
\begin{center}
\begin{spacing}{1.3}
\begin{tabular}{r|r|rr|rr} \hline\hline
\multicolumn{1}{c|}{No.} &
\multicolumn{1}{c|}{Normal} & \multicolumn{4}{c}{Localized Modes} \\ \hline
 & \multicolumn{1}{c|}{$\nu$} & \multicolumn{1}{c}{$\tilde{\nu}$} &  & \multicolumn{1}{c}{\mbox{L-VCI}-S} \\ \hline 
1 & 607 & 610 & 3 & 607 & 0 \\
2 & 614 & 610 & -3 & 614 & 0 \\ \cline{3-6}
3 & 736 & 788 & 52 & 736 & 0 \\
4 & 758 & 788 & 30 & 757 & 0 \\
5 & 854 & 824 & -30 & 853 & 0 \\
6 & 876 & 824 & -52 & 876 & 0 \\\cline{3-6}
7 & 879 & 879 & 0 & 879 & 0 \\
8 & 888 & 888 & 0 & 888 & 0 \\ \cline{3-6}
9 & 1012 & 1099 & 87 & 1012 & 0 \\
10 & 1063 & 1099 & 35 & 1062 & -2 \\
11 & 1089 & 1135 & 46 & 1087 & -2 \\
12 & 1161 & 1135 & -26 & 1158 & -3 \\
13 & 1218 & 1183 & -34 & 1217 & 0 \\
14 & 1291 & 1183 & -108 & 1297 & 6 \\ \cline{3-6}
15 & 1418 & 1438 & 20 & 1419 & 0 \\
16 & 1523 & 1548 & 25 & 1523 & 0 \\
17 & 1593 & 1548 & -45 & 1593 & 0 \\ \cline{3-6}
18 & 3255 & 3265 & 10 & 3255 & 0 \\
19 & 3266 & 3265 & -1 & 3265 & 0 \\
20 & 3286 & 3284 & -1 & 3286 & 0 \\
21 & 3292 & 3284 & -8 & 3292 & 0 \\ \hline
& & MAD & 29 &  & 1 \\
& & MAX & 108 &  & 6 \\ \hline\hline
\end{tabular}
\end{spacing}
\end{center}
\end{table}

\begin{figure}
\caption{Strengths of the two-mode couplings, $C(i,j)$, for the normal and localized modes (central figure) and total coupling of the $i$-th normal 
and localized mode with all other modes (bar plots), $C(i)$, for furan. The lower right part refers to normal modes, whereas the upper left 
part refers to localized modes. See Section~\ref{sec:ethene_harmonic} for further details.}
\label{fig:furan_nmvslm}
\begin{center}
\includegraphics[width=\textwidth]{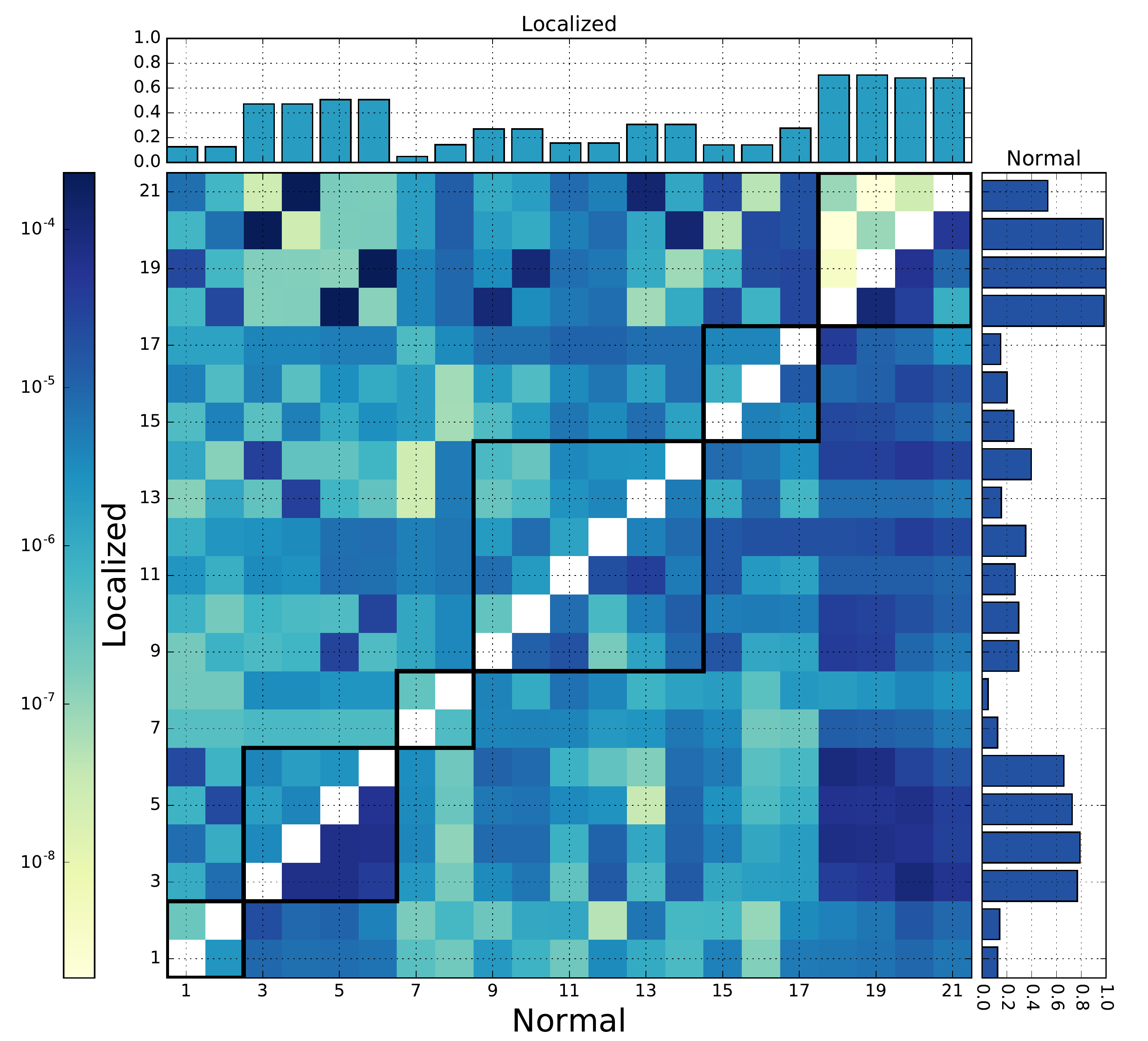}
\end{center}
\end{figure}

Also for furan, we compare the two-mode couplings in normal-mode coordinates and in localized-mode coordinates. These are visualized in
Fig.~\ref{fig:furan_nmvslm}. Similarly to the ethene case, in localized-mode coordinates the mutual couplings are weaker, which is especially 
pronounced for the C--H out-of-plane bending vibrations (modes 3--6), and for the C--H stretching vibrations (modes 18--21). This also shows
up in the total couplings shown in the bar charts, which are visibly smaller for the localized modes of these subsets compared to corresponding 
normal modes. In general, the localized modes are weaker coupled, with some single strong couplings, whereas for the normal modes one finds 
rather uniformly strong couplings. This aspect is especially noticeable for the well-localized subsets of modes 3--6 and modes 18--21.

\subsection{Convergence of the $n$-mode expansion and hybrid PES}
\label{sec:furan_potentials}

We have performed (L-)VCI-SDTQ5 calculations for different potential energy surfaces expanded in
normal-mode coordinates and in localized-mode coordinates. These will be introduced in the following. The resulting
fundamental vibrational frequencies are compared to the experimental results in Table~\ref{tab:furan_anh}.

\begin{table}
\caption{(L-)VCI-SDTQ56 fundamental vibrational frequencies calculated for furan with different approximations
for the PES (CC -- contributions obtained with CCSD(T)-F12x, DFT -- contributions obtained with DFT/BP/def2-TZVP; 
$^{(n)}$ -- order in the $n$-mode expansion, $^{(2{\rm h})}$~-- two-mode potentials obtained within the harmonic approximation, 
$^{(2{\rm s})}$ -- two-mode potentials in which the contributions of modes belonging to the same subset have been removed;
See text for details). In each column, the frequencies are given together with their deviations from the experimental 
reference value as well as the mean average deviation (MAD) and their maximum absolute deviation (MAX). 
All frequencies given in~\rcm.}
\label{tab:furan_anh}
\begin{center}
\begin{spacing}{1.3}
\hspace*{-0.5cm}
\begin{tabular}{r|rr|rr|rr|rr|rr|rr|r} \hline\hline
\multicolumn{1}{c|}{No.} & \multicolumn{6}{c|}{Normal Modes} & \multicolumn{6}{c|}{Localized Modes}  & Exp.$^{\rm a}$ \\ \cline{2-13}
 & \multicolumn{2}{c|}{$V^{(1)}_{\rm CC}$}  & \multicolumn{2}{c|}{$V^{(1)}_{\rm CC} + V^{(2)}_{\rm DFT}$} &
\multicolumn{2}{c|}{$V^{(1,2,3)}_{\rm CC}$}      & \multicolumn{2}{c|}{$V^{(1,2{\rm h})}_{\rm CC}$}   & \multicolumn{2}{c|}{$V^{(1,2{\rm h})}_{\rm CC} + V^{(2{\rm s})}_{\rm DFT}$}   & \multicolumn{2}{c|}{$V^{(1)}_{\rm CC} + V^{(2)}_{\rm DFT}$} & \\ \hline
1 & 608 & 7 & 606 & 6 & 596 & -3 & 609 & 9 & 602 & 2 & 601 & 1 & 600 \\
2 & 614 & 9 & 610 & 7 & 600 & -3 & 615 & 12 & 604 & 1 & 605 & 2 & 603 \\
3 & 786 & 62 & 759 & 37 & 726 & 4 & 836 & 114 & 711 & -11 & 724 & 3 & 722 \\
4 & 791 & 47 & 781 & 36 & 747 & 3 & 855 & 110 & 738 & -7 & 744 & -1 & 745 \\
5 & 880 & 57 & 872 & 35 & 835 & -2 & 880 & 42 & 830 & -8 & 823 & -15 & 838 \\
6 & 881 & 29 & 874 & 10 & 860 & -4 & 888 & 24 & 853 & -11 & 851 & -13 & 864 \\
7 & 888 & 14 & 878 & 8 & 868 & -2 & 941 & 70 & 874 & 4 & 874 & 3 & 870 \\
8 & 911 & 29 & 888 & 15 & 878 & 5 & 962 & 89 & 882 & 9 & 881 & 8 & 873 \\
9 & 1020 & 26 & 1001 & 6 & 994 & -1 & 1035 & 40 & 990 & -5 & 994 & -1 & 995 \\
10 & 1069 & 27 & 1053 & 10 & 1041 & -2 & 1075 & 32 & 1039 & -3 & 1042 & -1 & 1043 \\
11 & 1086 & 9 & 1071 & 4 & 1067 & 0 & 1085 & 18 & 1064 & -3 & 1077 & 10 & 1067 \\
12 & 1164 & 21 & 1146 & 6 & 1139 & -1 & 1174 & 34 & 1138 & -2 & 1144 & 3 & 1140 \\
13 & 1223 & 44 & 1189 & 8 & 1183 & 2 & 1220 & 39 & 1195 & 14 & 1179 & -2 & 1181 \\
14 & 1300 & 45 & 1274 & 7 & 1265 & -1 & 1314 & 48 & 1275 & 8 & 1256 & -11 & 1267 \\
15 & 1421 & 27 & 1395 & 10 & 1383 & -2 & 1421 & 37 & 1390 & 6 & 1394 & 10 & 1385 \\
16 & 1523 & 38 & 1495 & 5 & 1495 & 5 & 1520 & 30 & 1492 & 2 & 1485 & -5 & 1491 \\
17 & 1599 & 36 & 1566 & 9 & 1557 & 0 & 1589 & 32 & 1561 & 4 & 1563 & 6 & 1558 \\
18 & 3259 & 140 & 3066 & -64 & 3103 & -27 & 3144 & 14 & 3120 & -10 & 3120 & -10 & 3130 \\
19 & 3267 & 138 & 3095 & -45 & 3124 & -16 & 3155 & 15 & 3126 & -14 & 3129 & -11 & 3140 \\
20 & 3308 & 153 & 3094 & -66 & 3143 & -17 & 3175 & 14 & 3155 & -6 & 3155 & -6 & 3161 \\
21 & 3337 & 162 & 3156 & -13 & 3147 & -22 & 3182 & 12 & 3173 & 3 & 3175 & 6 & 3169 \\
\hline
& \multicolumn{1}{c}{MAD} & 53 &  & 19 &  & 6 &  & 40 &  & 6 &  & 6 \\
& \multicolumn{1}{c}{MAX} & 162 &  & 66 &  & 27 &  & 114 &  & 14 &  & 15 \\ \hline\hline
\multicolumn{14}{l}{$^{\rm a}$ -- experimental values references in Ref.~\cite{billes_vibrational_2004}} 
\end{tabular}
\end{spacing}
\end{center}
\end{table}

When expanding the PES in normal-mode coordinates, VCI calculations with an accurate PES including up to three-mode potential, 
denoted $V^{(1,2,3)}_{\rm CC}$, yield fundamental transitions that are on average 6~\rcm\ off from the experiment. The largest deviations
of up to 27~\rcm are observed for the C--H stretching vibrations. To construct such a PES directly, neglecting symmetry and automated fitting 
procedures and assuming a 16-point grid for each mode, requires performing 336 CCSD(T)-F12b/cc-pVTZ-F12 single-point calculations for 
the one-mode potential, and in total 5\ 501\ 776 CCSD(T)-F12a/cc-pVDZ-F12 single-point calculations for the two- and three-mode terms.

For an approximation that is computationally less demanding, we have decided to use a hybrid PES, in which CCSD(T)-F12b/cc-pVTZ-F12
is used for the one-mode potentials, while DFT/BP/def2-TZVP calculations were employed for the two-mode terms and the three-mode
potentials are neglected. This hybrid PES is denoted $V^{(1)}_{\rm CC} + V^{(2)}_{\rm DFT}$. If normal-mode coordinates are used, 
VCI-SDTQ5 calculations with such a PES give fundamental frequencies that deviate on average 19~\rcm\ and at most 66~\rcm\ from 
the reference values. 

In contrast, if localized-mode coordinates are used, such calculations yield frequencies that are on average 6~\rcm\ and at most 15~\rcm\ 
off from the experimental results. Thus, the error is reduced by a factor of 3--4 compared to normal modes. This reduction of the error is
especially pronounced for the modes that are well localizable, in particular for modes 3--6 and modes 18--21 (cf. Fig.~\ref{fig:furan_modes}).
Note that in terms of localized-mode coordinates, such a hybrid PES can provide results that are as accurate as the full CC PES including 
three-mode potentials expanded in normal-mode coordinates.

As a further approximation, the off-diagonal elements of Hessian in the basis of localized modes are used to calculate harmonic two-mode potentials.
These can be used to replace the full anharmonic two-mode potential for the modes belonging to the same subset in the localization. This
simplified hybrid PES is denoted as $V^{(1,2{\rm h})}_{\rm CC} + V^{(2s)}_{\rm DFT}$. This way, 32 two-mode potentials (corresponding to
8\,192 single-point calculations) can be omitted. Such an approximation results in an MAD of the fundamental frequencies of 6 \rcm\ and
a maximum deviation of 14 \rcm\, which is identical to the accuracy obtained when included all DFT two-mode potentials.

Finally, as a low-cost approximation we use only the harmonic two-mode potentials along with anharmonic one-mode potentials, both
expanded in localized-mode coordinates, which requires only 336 single-point CCSD(T)-F12b/cc-pVTZ-F12 calculations. This PES is 
denoted as $V^{(1,2{\rm h})}_{\rm CC}$. This computationally very cheap approach results in an MAD of 40~\rcm, and a maximal deviation 
of 114~\rcm. It is noteworthy that the C--H stretching vibrations are very well reproduced and are only about 14~\rcm\ off from the 
experimental values. This is a closer agreement than for the full $V^{(1,2,3)}_{\rm CC}$ PES expanded in normal-mode coordinates.
If normal-mode coordinates are used, we can obtain the anharmonic one-mode potentials ($V^{(1)}_{\rm CC}$) with the same
computational effort. As in the ethene example, this results in rather large discrepancies for the C--H stretching modes, whereas for 
the other vibrations, the deviations are at the similar as for localized modes. All in all, the MAD raises to 53~\rcm, and the maximal deviation 
to 162~\rcm\ when applying this low-cost approximation with normal-mode coordinates instead of localized-mode coordinates.

In summary, for furan similar trends as in the ethene example are observed. In particular, for the modes that can be well localized,
such as the C--H stretching and (to a smaller extent) bending vibrations, the convergence of the $n$-mode expansion is significantly
faster in localized-mode coordinates. This makes it possible to employ computationally cheaper hybrid potential energy surfaces. If
localized-mode coordinates are used, a hybrid $V^{(1)}_{\rm CC} + V^{(2)}_{\rm DFT}$ yields fundamental vibrational frequencies
of the same accuracy as a $V^{(1,2,3)}_{\rm CC}$ expanded in normal-mode coordinates.
Moreover, in localized-mode coordinates it becomes possible to neglect some of the two-mode potentials by replacing them with 
the harmonic counterparts arising from the localization without loss of accuracy. Finally, localized-mode coordinates are better
suited for devising low-cost approximations that require only the calculation of anharmonic one-mode potentials and that can 
provide a first approximation of anharmonic corrections.

\section{Conclusions and summary}
\label{sec:summary}

In this paper, we have presented L-VSCF/L-VCI calculations performed for all fundamental vibrations in two prototypical 
small molecules, ethene and furan. We have investigated to what extent employing localized-mode coordinates instead of
the conventionally used normal-mode coordinates can be beneficial with respect to the main bottlenecks of such anharmonic 
vibrational calculations. 

Concerning the convergence of the $n$-mode expansion, we observe a significantly faster convergence in localized-mode 
coordinates. While for ethene, the inclusion of up to four mode potentials is required to reduce the mean average deviation
from the experimental fundamental frequencies below 10~\rcm\ when using normal-mode coordinates, with localized-mode
coordinates the same level of accuracy can already be reached with only up to three-mode potentials. Similarly, for furan
it is possible to approximate the two-mode potentials with lower-level DFT calculations when using localized-mode coordinates,
while still achieving the same level of accuracy as with a high-level CCSD(T)-F12x PES including up to three-mode potentials
in terms of normal-mode coordinates.  Moreover, we observe that the two-mode couplings are significantly reduced in localized-mode
coordinates, in particular within the subsets of modes that are used in the localization procedure. This can be exploited to
neglect selected two-mode potentials or to replace them by their harmonic counterparts, without losing accuracy for the
fundamental frequencies. In combination, these advantageous features of localized-mode coordinates can vastly reduce 
the number of single-point calculations needed to construct the PES and thus help to alleviate this computational bottleneck.

In addition, we find that the convergence with respect to the VCI excitation space proceeds more smoothly when using
localized-mode coordinates. This could be exploited to reduce the computational effort required for this step, for instance
by making schemes for the selection of the relevant excitations\cite{rauhut_configuration_2007, scribano_iterative_2008, 
carbonniere_vci-p_2010,knig_automatic_2015} more efficient. Moreover, as the error with small VCI excitation spaces
is significantly reduced when using localized-mode coordinates, it becomes possible to devise low-cost models that can
give a first estimate of anharmonic frequency corrections. For our two test cases, we find that L-VCI-S with only anharmonic
one-mode potentials and harmonic two-mode potentials can provide such a qualitatively correct estimate for all fundamental
vibrations, whereas with normal-mode coordinates, at least anharmonic two-mode potentials in combination with a significantly
larger excitation space would be required.

While similar benefits can also be achieved with optimized coordinates \cite{thompson_optimization_1982,moiseyev_scf_1983,
yagi_optimized_2012, thomsen_optimized_2014}, localized-mode coordinates offer the advantage that they can be constructed 
\textit{a priori}, i.e., no anharmonic PES is required for their construction. Moreover, compared to other local 
coordinates\cite{low_calculation_1999, salmi_calculation_2009, wang_towards_2010} or local (curvilinear) internal 
coordinates \cite{yurchenko_theoretical_2007,scribano_fast_2010} that could be employed to 
achieve such benefits in anharmonic vibrational calculations, the rigorously-defined localized modes used here do not 
require the manual \textit{ad hoc} construction of vibrational coordinates. Instead, only a chemically meaningful assignment
of the normal modes to subsets is required as prerequisite for the localization procedure. The effect that the choice of
these subsets has on the accuracy of the anharmonic vibrational frequencies will be subject of our future work.

Thus, our results demonstrate that normal modes may not always be optimal for anharmonic vibrational calculations.
In particular for well-localizable vibrations, such as C--H stretching and bending modes, localized modes are in general a 
better choice. The benefits of localized modes will become more pronounced with increasing size of the molecule, for which
a better localization will be possible and for which further criteria, such as distance cut-offs,\cite{cheng_efficient_2014} can be 
applied. Finally, the strategies presented here are not limited to the calculation of fundamental frequencies, but will also be
applicable when targeting overtones and combination bands. This will be addressed in our future work.

\section*{Acknowledgments}

CRJ is grateful to the Deutsche Forschungsgemeinschaft (DFG) for funding via grant JA 2329-2/1.
PTP would like to thank Prof.~Wim Klopper for his kind hospitality in his group at the Karlsruhe Institute of Technology (KIT).

\end{document}